\documentclass[aps,twocolumn,prl,showpacs,amsmath,superscriptaddress,amssymb,showkeys,10pt]{revtex4-1}
\usepackage{graphicx}
\usepackage{epstopdf}
\usepackage{braket}
\usepackage{hyperref}
\allowdisplaybreaks

\begin{document}

\title{Telling emissions apart in a multiphoton resonance: visualizing a conditional evolution}

\author{Th. K. Mavrogordatos}
\email[Email address(es): ]{themis.mavrogordatos@icfo.eu;  themis.mavrogordatos@fysik.su.se}
\affiliation{ICFO -- Institut de Ci\`{e}ncies Fot\`{o}niques, The Barcelona Institute of Science and Technology, 08860 Castelldefels (Barcelona), Spain}
\affiliation{Department of Physics, Stockholm University, SE-106 91 Stockholm, Sweden}

\date{\today}

\begin{abstract}
We find that the phase-space representation of the electromagnetic field inside a driven cavity strongly coupled to a two-level atom can be employed to distinguish photon emissions along a ladder of dressed states sustaining a two-photon resonance. The emissions are told apart by means of the different quantum beats generated by the conditional states they prepare. Sample quantum trajectories explicitly reveal the difference in the transient due to the initial condition, in a background set by the Jaynes-Cummings spectrum and revealed by the strong-coupling limit. Their ensemble-averaged evolution is tracked for a time period similar to that waited for the loss of a next photon as the maximum non-exclusive probability, indicated by the peak of the intensity correlation function.    
\end{abstract}


\maketitle

The formation of multiphoton resonances as a consequence of nonlinear light-matter interaction in the regime of non-perturbative quantum electrodynamics (QED) is a phenomenon of fundamental interest in modern optics since they dynamically probe the interface between classical and quantum physics. In the open driven Jaynes-Cummings (JC) model, these resonances arise as a result of the ``quantum jumpiness'' in the matter~\cite{CarmichaelTalk2000}, for a scattering process fundamentally affected by the inputs and outputs -- information is read off two channels linked to a system comprising a two-level atom strongly coupled to a resonant coherently driven cavity mode. The special role of quantum fluctuations in this model is captured by the breakdown of photon blockade by means of a dissipative quantum phase transition~\cite{Alsing1991, Kilin91, Sarabi2015, PhotonBlockade2015, Fink2017, Pietikainen2019}. 

Following the first observation of the two-photon resonance reported in 2008~\cite{Schuster2008}, the recent experiment conducted by Najer and collaborators~\cite{Najer2019} identifies photon bunching in the correlation function of the transmitted light [as depicted in Figs. 4(b) and S1 of their report]. Preferential scattering of Fock states $\ket{n}$ with $n\geq 2$ occurred, alongside fast oscillations at twice the frequency of the light-matter coupling strength when driving the two-photon transition in a system comprising a gated quantum dot strongly coupled to an optical microcavity. Another recent cavity QED experiment realized by Hamsen and coworkers also focuses on the JC two-photon transition~\cite{Hamsen2017}; these results also evince photon bunching through the second-order correlation function of the forwards scattered photons. The function peaks at a finite time delay, violating the Schwartz inequality~\footnote{The correlation function of a classical intensity is constrained by the inequality $|g^{(2)}(\tau)-1| \leq g^{(2)}(0)-1 \geq 0$, as shown in Ref.~\cite{Rice1988}.} -- an unequivocally quantum feature~\cite{Mielke1998, Foster2000B}. While scattering from a two-photon transition is commonly associated with bunching, Shamailov and collaborators reported in 2010 that both bunching and antibunching occur depending on the strength of the driving in the strong-coupling limit~\cite{Shamailov2010}. They also identified a quantum beat in the photon correlation arising when the two intermediate states occupied through the cascaded decay are prepared as a superposition after a single photon is emitted from the cavity. The quantum beat oscillates at the frequency difference between the excited couplet states, directly revealing the strong light-matter coupling strength in excess of the level widths. This difference, which for the first excited coupled is equal to twice the light-matter coupling strength, can be orders of magnitude higher than the semiclassical Rabi splitting. 

What can such a characteristic separation of time scales tell us about the JC conditional evolution in a regime where a photon emission is noticeable and recordable as an observable change~\cite{Foster2000A, Foster2000B, Walther2006, Dayan2008, Faraon2008, Kubanek2008, Faraon2010, Chang2014}? To address our question, in this Letter we demonstrate that the generation of the quantum beat is conditioned on the type of emission occurring after the attainment of steady state, down one of the two dissipation channels where the output information is read. We employ a perturbative method to calculate the Wigner function of the electromagnetic field state as a means to image the subsequent evolution of the intracavity field in a part of the transient where the {\it quasi}probability distribution may take negative values. We also show that one can infer the conditional state following a photon emission based on the fraction of trajectories in which the quantum beat prominently features, as a result of an interference indicating the excitation of the first couplet along the JC ladder -- the very same interference that makes the Wigner function turn negative. 

In the spirit of Ref.~\cite{Shamailov2010}, a four-level minimal model was employed in Ref.~\cite{Lledo2021} for the study of the intensity correlation function of the fluorescence emitted when maintaining a JC two-photon resonance, while the associated Wigner function of the cavity field was analytically determined in Ref.~\cite{Wigner2PB}. The results presented herein point to the link between cavity-state tomography in the presence of single-photon Fock states pioneered by Lutterbach and Davidovich~\cite{Lutterbach1997} and subsequently demonstrated by Nogues and collaborators~\cite{Nogues2000}, and the recent cavity QED experiments initiated by Hamsen and collaborators. To make the connection we operationally distinguish the transitions encountered in a two-photon resonance from the coherence properties that are read as a consequence of the related conditioning -- in fact, the same conditioning underlying the definition of a second-order correlation function. We should keep in mind that the intensity correlation function $g^{(2)}(t,t+\tau)$ of, say, a cavity field, is to be read as the product of the mean photon number at time $t$, and the mean photon number at time $t + \tau$, {\it subject to the condition} that a photon was detected from the field at time $t$~\cite{OpenQuantCh4}. In what follows, we consider the steady-state limit, with $g^{(2)}(\tau) \equiv \lim_{t\to\infty} g^{(2)}(t,t+\tau)$. This means that we are dealing with stationary fluctuations; all probabilities and averages are independent of $t$, while the correlation function only depends on the time difference $\tau$ (see Sec. 2.3.5 of Ref.~\cite{CarmichaelQO1}). 

Let us now start with locating the transitions in question among the dressed states the system dynamics is primarily constrained to follow in our description. The density matrix $\rho$ of the atom-cavity system, coupled to two independent reservoirs (external electromagnetic field) in the vacuum state, obeys the standard Lindblad master equation (ME)
\begin{equation}\label{eq:ME1}
\begin{aligned}
 \frac{d\rho}{dt}=&-i[\omega_0(\sigma_{+}\sigma_{-} + a^{\dagger}a)+g(a\sigma_{+}+a^{\dagger}\sigma_{-}),\rho]\\
 &-i[\varepsilon_d (a e^{i\omega_d t} + a^{\dagger}e^{-i\omega_d t}),\rho]\\
 &+\kappa (2 a \rho a^{\dagger} -a^{\dagger}a \rho - \rho a^{\dagger}a)\\
 &+\frac{\gamma}{2}(2\sigma_{-}\rho \sigma_{+} - \sigma_{+}\sigma_{-}\rho - \rho \sigma_{+}\sigma_{-}),
 \end{aligned}
\end{equation}
where $a$ ($a^{\dagger}$) are the annihilation (creation) operators for the cavity photons, $\sigma_{+}$ ($\sigma_{-}$) are the raising (lowering) operators for the two-level atom, $g$ is the dipole coupling strength, $2\kappa$ is the photon loss rate from the cavity (forwards scattering), and $\gamma$ is the spontaneous emission rate for the atom (side scattering) to modes other than the privileged cavity mode. The cavity is coherently driven with amplitude $\varepsilon_d$ and frequency $\omega_d$.
\begin{figure}
 \begin{center}
\includegraphics[width=0.3\textwidth]{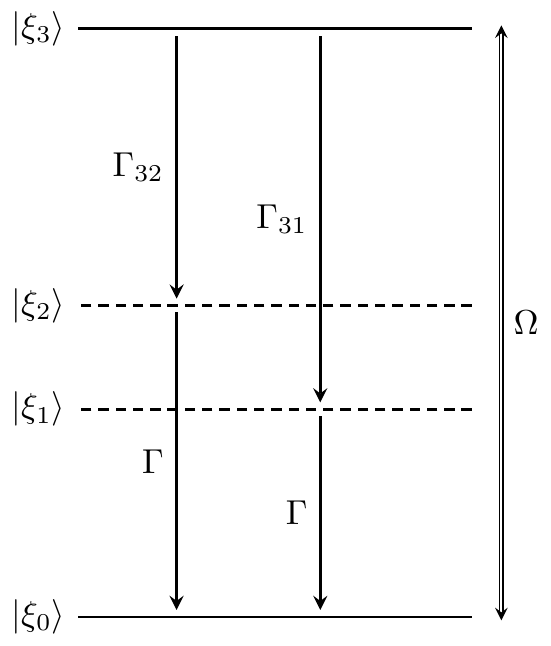}
\end{center}
\caption{{\it Schematic of the minimal model.} The displayed energy levels, given by Eqs~\eqref{eq:shifts}, are further dressed by the interaction with the drive. The effective two-photon Rabi frequency is $\Omega=2\sqrt2 \varepsilon_d^2/g$. The intermediate states $\ket{\xi_1}, \ket{\xi_2}$, marked by dashed lines, may be adiabatically eliminated but not discarded, since they are occupied in the cascaded decay~\cite{Shamailov2010}.}
\label{fig:4lmodel}
\end{figure}

The concept of photon blockade has been introduced by Imamo\u{g}lu and collaborators in close analogy with the effect of Coulomb blockade for quantum-well electrons~\cite{Imamoglu1997}. While the connection between the antibunching~\cite{Kimble1977, Diedrich1987, Rotter2008} of quasi-elastically scattered photons and one-photon blockade~\cite{Birnbaum2005, Lang2011} is transparent, the photon statistics associated with a two-photon resonance~\cite{Zubairy1980, Hughes2011, Reinhard2012, Munoz2014} reflects the dynamical nature~\cite{Fink2008} of the excitation bottleneck ``caused by detuning when one tries to drive monochromatic photons up an energy ladder of unequal step''~\cite{PhotonBlockade2015}. The requirement of two-photon blockade is expressed through $(3-\sqrt{6})g \gg \sqrt{2} (\kappa, \gamma/2)$, while the condition $(\sqrt{2}-1)g \gg \sqrt{2}(\kappa, \gamma/2)$ guarantees that the two-photon frequency $\omega_0-g/\sqrt{2}$ is sufficiently detuned from the one-photon resonance~\cite{Shamailov2010}. In the strong-coupling limit, the fine scale of the energy spectrum comprising the ground state and the JC couplets, $E_{n,\pm}=n\hbar \omega_0  \pm \sqrt{n} \hbar g$. In the regime of operation we are considering, with $g/\kappa=1000$ [we take our justification for the very strong coupling from Refs.~\cite{Bishop2009, yoshihara2017superconducting, Sett2022}\footnote{Provided of course that $g/\omega_0 <0.1$ so that the rotating-wave approximation holds.}], this spectrum admits perturbative corrections due to presence of the external drive; for the two-photon resonance, the expansion is made in powers of $(\varepsilon_d/g)^2 \ll 1$, forming the basis for a perturbative method after~\cite{GottfriedQM} holding for $(1/2)(\kappa + \gamma/2)<\varepsilon_d \ll g$. Tian and Carmichael point to saturation as a distinct consequence of the two-state behavior produced by photon blockade. To interpret the semiclassical Rabi splitting and the Mollow triplet spectrum~\cite{Mollow1969} arising when driving the vacuum Rabi resonance at significant excitation, they speak of {\it dressing of the dressed states}~\cite{Tian1992}, an effect which has been experimentally verified in circuit QED~\cite{Bishop2009}. For the purposes of our discussion we recast the ME~\eqref{eq:ME1} in an alternative form, concentrating on a certain part of the JC ladder for which the addition or subtraction of one or two quanta is recorded as an observable modification of the dynamics. We then truncate the Hilbert space to the first four dressed JC states between which transitions occur, as depicted in Fig.~\ref{fig:4lmodel} (after Fig. 2 of Ref.~\cite{Shamailov2010}). The set comprises the ground state, the first excited couplet and the lower state from the second excited couplet:
\begin{subequations}\label{eq:4levels}
\begin{align}
 &\ket{\xi_0}\equiv\ket{0,-}, \\
 &\ket{\xi_1}\equiv \frac{1}{\sqrt{2}} (\ket{1,-} - \ket{0,+}), \\
 & \ket{\xi_2}\equiv \frac{1}{\sqrt{2}} (\ket{1,-} + \ket{0,+}), \\
 &  \ket{\xi_3}\equiv \frac{1}{\sqrt{2}} (\ket{2,-} - \ket{1,+}). 
\end{align}
\end{subequations}
In this unperturbed form, with $\ket{n, \pm}\equiv \ket{n}\otimes\ket{\pm}$, $\ket{n}$ is the Fock state of the cavity field, while $\ket{+}, \ket{-}$ are the upper and lower states of the two-level atom, respectively. Under two-photon blockade operation, transitions to higher levels of the JC ladder are assumed far from resonance. Within the truncation we adopt, the photon annihilation operator and the atomic lowering operator read
\begin{subequations}\label{eq:jumpops}
\begin{align}
 &a=\frac{1}{\sqrt{2}}|\xi_0 \rangle \left(\langle \xi_1|+ \langle \xi_2|\right) + \left(\lambda_{+} |\xi_0 \rangle +  \lambda_{-} |\xi_2 \rangle \right) \langle \xi_3|,\label{eq:aop}\\
 &\sigma_{-}=-\frac{1}{\sqrt{2}}|\xi_0 \rangle \left(\langle \xi_1| - \langle \xi_2|\right)-\frac{1}{2}(|\xi_1 \rangle + |\xi_2 \rangle) \langle \xi_3|, \label{eq:sigmaop}
\end{align}
\end{subequations}
respectively, with $\lambda_{\pm}=(\sqrt{2}\pm 1)/2$. The corresponding jump operators are $J=(\sqrt{2\kappa}\, a, \sqrt{\gamma}\, \sigma_{-})$ for forwards and side emission, respectively. In what follows we also take $\gamma=2\kappa$ to simplify the expressions resulting from the application of the perturbative method.

Making the secular approximation~\cite{Cohen_Tannoudji_1977}, the strong-coupling limit of nonperturbative QED ($g \gg \kappa, \gamma/2$) leads to the following effective ME [see also Eq. (18) of Ref.~\cite{Shamailov2010}]
\begin{equation}\label{eq:ME2}
\begin{aligned}
  &\frac{d\rho}{dt}=\mathcal{L}\rho \equiv -i[\tilde{H}_{\rm eff},\rho]+\Gamma_{32} \mathcal{D}[|\xi_2\rangle \langle \xi_3|](\rho)\\
  &+ \Gamma_{31} \mathcal{D}[|\xi_1\rangle \langle \xi_3|](\rho)
  + \Gamma \mathcal{D}[|\xi_0\rangle \langle \xi_1|](\rho) + \Gamma \mathcal{D}[|\xi_0\rangle \langle \xi_2|](\rho),  
  \end{aligned}
\end{equation}
where
\begin{equation}
 \tilde{H}_{\rm eff}\equiv\sum_{k=0}^{3} \tilde{E}_{k} |\xi_k\rangle \langle \xi_k| + \hbar \Omega (e^{2i\omega_d t} |\xi_0\rangle \langle \xi_3| + e^{-2i\omega_d t} |\xi_3\rangle \langle \xi_0|).
\end{equation}
The shifted energy levels dressed by the drive are
\begin{subequations}\label{eq:shifts}
\begin{align}
 &\tilde{E}_0=E_0 + \hbar\delta_0(\varepsilon_d) = \hbar \sqrt{2} \varepsilon_d^2/g, \\
 & \tilde{E}_1=E_1 + \hbar\delta_1(\varepsilon_d) = \hbar \{\omega_0 -  g - [(20 + 19\sqrt{2})/7]\varepsilon_d^2/g\}, \\
 & \tilde{E}_2=E_2 + \hbar\delta_2(\varepsilon_d) = \hbar \{\omega_0 +  g +  [(20 - 19\sqrt{2})/7]\varepsilon_d^2/g\}, \\
 & \tilde{E}_3=E_3 + \hbar\delta_3(\varepsilon_d) =  \hbar (2\omega_0 - \sqrt{2}  g -  \sqrt{2}\, \varepsilon_d^2/g),
\end{align}
\end{subequations}
while the effective two-photon Rabi frequency is $\Omega=2\sqrt{2}\, \varepsilon_d^2/g$~\cite{Lledo2021}. In the effective ME~\eqref{eq:ME2}, we define $\mathcal{D}[X](\rho)\equiv X\rho X^{\dagger}-(1/2)\{X^{\dagger}X, \rho\}$, while the transition rates between the four states are given in the secular approximation as~\cite{CarmichaelQO2, Shamailov2010}
\begin{subequations}
\begin{align}
 &\Gamma_{31}\equiv\frac{\gamma}{4}+(\sqrt{2}+1)^2 \frac{\kappa}{2}=\frac{\gamma}{4}[1+(\sqrt{2}+1)^2], \label{eq:Gamma31} \\
 & \Gamma_{32}\equiv\frac{\gamma}{4}+(\sqrt{2}-1)^2 \frac{\kappa}{2}=\frac{\gamma}{4}[1+(\sqrt{2}-1)^2], \\
 & \Gamma\equiv\frac{\gamma}{2}+\kappa=\gamma,
\end{align}
\end{subequations}
The two-photon resonance must be excited with a drive frequency $\omega_d$ obeying $2\omega_d=(\tilde{E}_3-\tilde{E}_1)/\hbar=2\omega_0-\sqrt{2}g + \delta_3(\varepsilon_d)-\delta_0(\varepsilon_d)$, yielding $\Delta\omega_d\equiv \omega_d-\omega_0=-g/\sqrt{2} - \sqrt{2}\, \varepsilon_d^2/g$.

Let us next concentrate on the solution of the ME~\eqref{eq:ME2} within the secular approximation in a frame rotating with the drive frequency. The steady state is determined by means of detailed balance for the driven transition as
\begin{equation}
 \rho_{\rm ss}=\sum_{i=0}^{3} p_i |\xi_i \rangle \langle \xi_i| + i\frac{\gamma}{\Omega}p_3(|\xi_0 \rangle \langle \xi_3|-|\xi_3 \rangle \langle \xi_0|),
\end{equation}
where the steady-state occupation probabilities of the shifted energy levels comprising the minimal model are $p_0=1-3p_3$, $p_1=(\Gamma_{31}/\Gamma)p_3$, $p_2=(\Gamma_{32}/\Gamma)p_3$, with $p_3=\Omega^2/(\gamma^2 + 4\Omega^2)$. The two states $|\xi_1 \rangle$ and $|\xi_2 \rangle$ cannot be adiabatically eliminated since they mediate the cascaded decay. They are responsible for the generation of the quantum beat which only contributes to the transient as $\rho_{12}(\tau)=\rho_{21}^{*}(\tau)=\rho_{12}(0)\,e^{-\gamma \tau} e^{i \nu \tau}$, where $\nu=2g + \delta_2-\delta_1$ is the beat frequency capturing the splitting of the first excited couplet states. It is important to note that, within the secular approximation, $\rho_{12}(\tau)$ satisfies an equation of motion which is uncoupled to the rest of the matrix elements. For the time-evolving occupation probability of the upper level we find
\begin{equation}
\begin{aligned}
\rho_{33}(\tau)=&c_1 e^{-2\gamma\tau}-c_2 e^{-\gamma \tau} p_3[\gamma \sin(2\Omega \tau)-2\Omega \cos(2\Omega \tau)]\\
&+p_3\left[1-\frac{\gamma}{\Omega} e^{-\gamma \tau}\sin(2\Omega \tau)\right].
 \end{aligned}
\end{equation}
We note here that any post-steady-state photon emission we have $\rho_{33}(0)=0$ [see the expression of the jump operators in Eqs.~\eqref{eq:jumpops}], which gives $c_1=-p_3 (1+2c_2)$, while the coefficient $c_2$ has been defined as the effective conditional ``inversion'' of the two-photon process, given by the difference $\rho_{33}-\rho_{00}$ following the ``first'' detectable quantum event. Now, for the sum of probabilities to find the JC ``molecule'' in the intermediate states we obtain
\begin{equation}
\begin{aligned}
 &\rho_{11}(\tau) + \rho_{22}(\tau)= -2c_1 e^{-2\gamma \tau} \\ 
 &+2 c_2 e^{-\gamma \tau}p_3 \left[\left(\frac{\gamma}{\Omega}\right)^2 \cos(2\Omega \tau)+ \frac{\gamma}{\Omega}\sin(2\Omega \tau)\right]\\
 & +p_3\left[2-\left(\frac{\gamma}{\Omega}\right)^2 e^{-\gamma \tau}\cos(2\Omega \tau)\right],
 \end{aligned}
\end{equation}
while from the preservation of the trace we get $\rho_{00}(\tau)=1-[\rho_{11}(\tau) + \rho_{22}(\tau) + \rho_{33}(\tau)]$. Finally, for the two off-diagonal elements $\rho_{03}=\rho_{30}^*$, determining the ``polarization'' of the two-photon transition, we find ${\rm Re} [\rho_{03}(\tau)]=0$ and
\begin{equation}
\begin{aligned}
 2 {\rm Im} [\rho_{03}(\tau)]=&2\frac{\gamma}{\Omega}p_3 - e^{-\gamma \tau}\Big[c_2 \sin(2 \Omega \tau)\\
 &+ p_3 \left(\frac{\gamma}{\Omega}\right)^2 \sin(2 \Omega \tau) +2 p_3\frac{\gamma}{\Omega}\cos(2\Omega \tau) \Big].
 \end{aligned}
\end{equation}
The above expression shows that, for a saturated transition [$(\gamma/\Omega) \ll 1$], the variation of the ``polarization'' is primarily determined by the conditional ``inversion'' after a photon is emitted. 

\begin{figure*}
\begin{center}
\includegraphics[width=\textwidth]{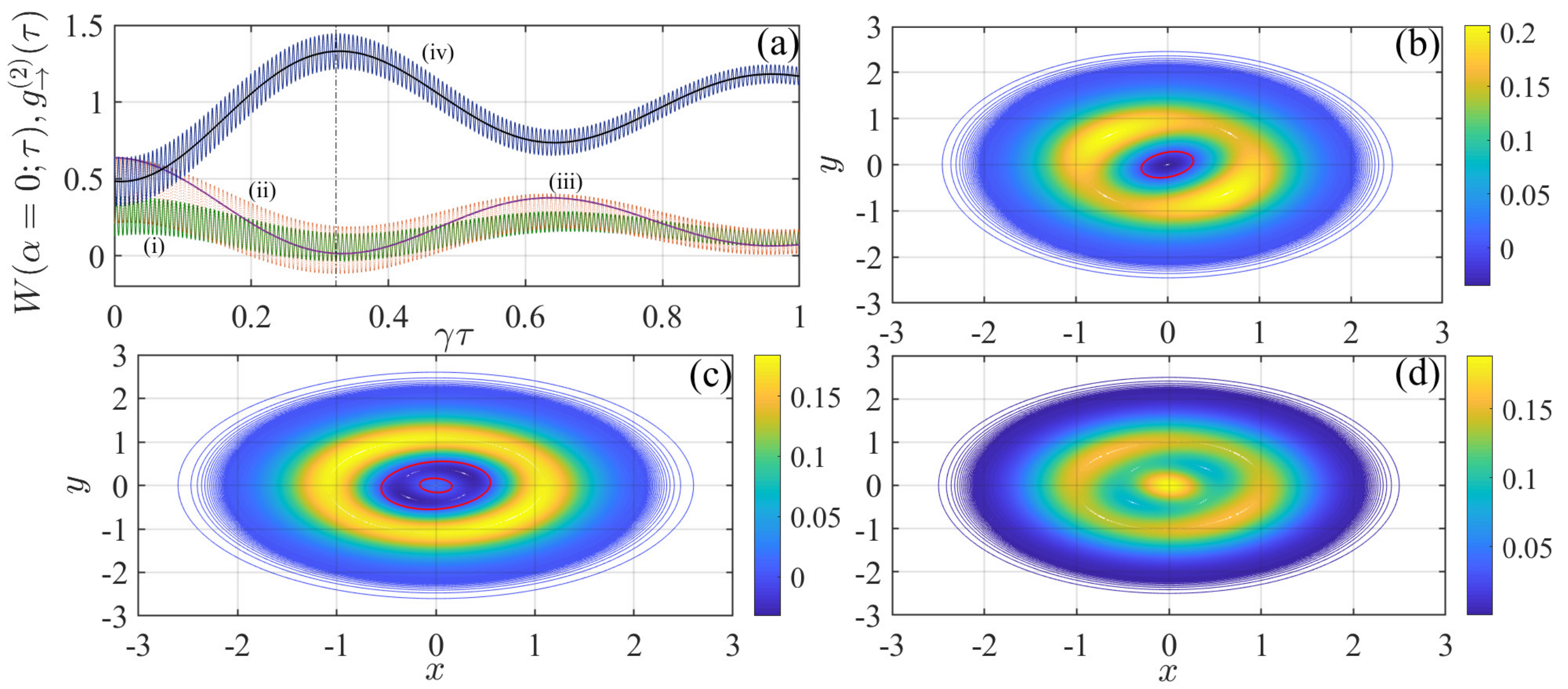}
\end{center}
\caption{{\it Conditional evolution of the field state in phase space.} {\bf (a)} The Wigner function of the intracavity field at the phase-space origin, $W(x=y=0;\tau)$, is plotted from Eq.~\eqref{eq:W0origin} for an initial single-photon emission in the forwards direction [thin solid green line (i)], a photon initially scattered off the sides of the cavity [dotted orange line (ii)] and a two-photon emission [thick solid purple line (iii)]. The line in blue, marked as (iv), depicts the intensity correlation function of the transmitted light, $g^{(2)}_{\rightarrow}(\tau)$, obtained from the analytical expression detailed in Eqs. (44--49) of Ref.~\cite{Shamailov2010}, while the average over the quantum beat is plotted by the solid black intersecting line. The vertical dash-dotted line indicates the time delay of maximum correlation, $\gamma \tau_{\rm max}=0.3235$. {\bf (b)} Wigner function $W(x+iy;\tau_b)$, following a {\it single-photon emission} in transmission, plotted for $\gamma \tau_b=0.3297$, the time when the corresponding function $W(x=y=0;\tau)$ [thin solid green line in (a)] reaches its global minimum value. The red contour marks the region inside which the Wigner function takes negative values. {\bf (c)} $W(x+iy;\tau_c)$, following a {\it two-photon emission} in transmission, plotted for $\gamma \tau_c=0.3327$, the time when $W(x=y=0;\tau)$, obtained for an initial side-scattering emission [dotted orange line in (a)],  reaches its global minimum value. The two contours in red mark the boundaries of the ring inside which the Wigner function is negative. {\bf (d)} $W(x+iy;\tau_d)$, following a {\it single-photon scattering} off the sides of the cavity, plotted for $\gamma \tau_d=0.3390$, the time when $W(x=y=0;\tau)$ [dotted orange line in (a)] reaches its first local maximum past its global minimum value. The contour plots in Figs.~\ref{fig:Wigfuncs}(b)--(d) are evaluated from Eq.~\eqref{eq:Wtau}. We take $p_3=0.2475$, $g/\gamma=500$ and $\gamma=2\kappa$.}
\label{fig:Wigfuncs}
\end{figure*}

The Wigner function $W(\tau)\equiv W(x+iy;\tau)$ of the intracavity field (with complex variables $\alpha=x + iy$, $\alpha^{*}=x-iy$) in the transient is defined as 
\begin{equation}
\begin{aligned}
&W(\tau) \equiv \frac{1}{\pi^2} \int d^2 z\,{\rm tr} \left( \rho_c(\tau) e^{iz^{*}a^{\dagger}+iza} \right) e^{-iz^{*}\alpha^{*}} e^{-iz\alpha},
\end{aligned}
\end{equation}
where $\rho_c(\tau) \equiv \braket{+|\rho(\tau)|+} + \braket{-|\rho(\tau)|-}$ is the reduced density matrix of the cavity field. From the solution of the ME~\eqref{eq:ME2} governing the evolution of matrix elements in the dressed-state basis, we may determine the time-varying coefficients of the various Fock-state contributions, both diagonal and off-diagonal, reading  
\begin{subequations}
 \begin{align}
d_0(\tau)&=\rho_{00}(\tau) + \frac{1}{2}[\rho_{11}(\tau)+ \rho_{22}(\tau)-\rho_{12}(\tau)-\rho_{21}(\tau)], \\
d_1(\tau)&=\frac{1}{2}[\rho_{11}(\tau)+ \rho_{22}(\tau)+ \frac{1}{2}[\rho_{12}(\tau)+ \rho_{21}(\tau)], \\
d_2(\tau)&=\frac{1}{2}\rho_{33}(\tau),\\
d_3(\tau)&=\frac{1}{\sqrt{2}}{\rm Im}[\rho_{03}(\tau)].
 \end{align}
\end{subequations}
Bringing the pieces together we find
\begin{equation}\label{eq:Wtau}
\begin{aligned}
 &W(x+iy;\tau)=\frac{2}{\pi}e^{-2|\alpha|^2}\Big\{d_0(\tau) - d_1(\tau) L_1(4|\alpha|^2)\\
 &+ d_2(\tau) L_2(4|\alpha|^2) +i2\sqrt{2}\,d_3(\tau)[\alpha^2-(\alpha^{*})^2]\Big\},
 \end{aligned}
\end{equation}
where $L_1$ and $L_2$ are the Laguerre polynomials of the first and second degree, respectively. All evolutions conditioned on different initial density matrices relax to the same steady state with a positive Wigner {\it quasi}probability distribution ($\alpha=x+iy$)~\cite{Wigner2PB},
\begin{equation}\label{eq:Wss}
\begin{aligned}
 &W_{\rm ss}(x+iy)=\frac{2}{\pi}e^{-2|\alpha|^2}\Big\{(1-2p_3) -  \frac{3}{2}p_3 L_1(4|\alpha|^2)\\
 &+ \frac{1}{2}p_3 L_2(4|\alpha|^2) +i\,2\frac{\gamma}{\Omega}p_3 [\alpha^2-(\alpha^{*})^2]\Big\}.
 \end{aligned}
\end{equation}
The above expression yields the phase-space representation of the electromagnetic field with an underlying dependence on a single parameter, the steady-state excitation probability of the upper level, $p_3$ [with $0\leq p_3 <0.25$, $\gamma/\Omega=\sqrt{(1-4p_3)/p_3}$]. It evinces the development of bimodality along the line $y=-x$ with increasing excitation. The numerical solution of the ME~\eqref{eq:ME1} via exact diagonalization of the Liouvillian shows a developing bimodality for $p_3 \gtrsim 0.13$, while for $p_3>0.24$, the two peak positions progressively divert from the line $y=-x$, and a single broad peak is formed~\cite{Wigner2PB}. This instance marks the borders of the region where perturbation theory is valid; for increasing $\varepsilon_d/g$ transitions occur to dressed states higher up in the ladder, with unequal detunings~\cite{Tian1992}. It is also worth remarking that if $\ket{\xi_5}\equiv \frac{1}{\sqrt{2}} (\ket{2,-} + \ket{1,+})$ is instead selected as the upper level of the two-photon transition (for a minimal model comprising five levels) then the perturbative treatment predicts a symmetric ``dressed'' detuning, $\Delta\omega_d=g/\sqrt{2} + \sqrt{2}\, \varepsilon_d^2/g$, while the Rabi frequency $\Omega=2\sqrt{2}\varepsilon_d^2/g$ remains unchanged; bimodality unfolds along the line $y=x$. 

We are now in position to discuss the impact of a single-photon or a two-photon emission to the transient following the corresponding perturbation to the steady state. We focus on three possible events: a single-photon emission is recorded either through forwards or side scattering, or a two-photon transition occurs following one of the two paths comprising the minimal model. We consider the first possibility, where emission of one forwards-scattered photon prepares the system in the state
\begin{equation}\label{eq:rhocondit}
\begin{aligned}
 \rho_{\rm cond}&=\frac{2}{5} |\xi_0\rangle \langle \xi_0| + \frac{3}{5} |\psi_{\rm s\,1}\rangle \langle \psi_{\rm s,\,1}|,
 \end{aligned}
\end{equation}
with 
\begin{equation}\label{eq:beat1}
 \ket{\psi_{\rm s,\,1}}=\sqrt{\frac{2}{3}}\left(\lambda_{+}\ket{\xi_1} +  \lambda_{-}\ket{\xi_2}\right).
\end{equation}
The record of a single-photon detector with jump super-operator $\mathcal{J}\rho=2\kappa a^{\dagger}\rho a$ sets the initial conditions for the matrix-element equations of motion to $c_2=\rho_{33}(0)-\rho_{00}(0)=-2/5$, $\rho_{11}(0) + \rho_{22}(0)=3/5$ and $\rho_{12}(0)=1/10$. The quantum beat is present and has partial visibility, since the expansion coefficients $\lambda_{\pm}$ in Eq.~\eqref{eq:beat1} are unequal. Evolving this state under the action of the Liouvillian $\mathcal{L}$ in Eq.~\eqref{eq:ME2} yields the second-order correlation function for the forwards-scattered photons discussed in Sec. 3.3 of Ref.~\cite{Shamailov2010}.

Moving on to the second possibility, the conditioning is on the first side-scattered emission corresponding to the jump super-operator $\mathcal{J}\rho=\gamma \sigma_{+}\rho \sigma_{-}$. From the expression of the atomic lowering operator of Eq.~\eqref{eq:sigmaop} we obtain
\begin{equation}\label{eq:beat2}
 \ket{\psi_{\rm s,\,2}}=\frac{1}{\sqrt{2}} (\ket{\xi_1}+\ket{\xi_2}),
\end{equation}
while the following mixed state is prepared:
\begin{equation}\label{eq:mixedspon}
 \rho_{\rm cond}(0)=\frac{2}{3}|\xi_0\rangle \langle \xi_0| + \frac{1}{3}|\psi_{\rm s,\,2} \rangle \langle \psi_{\rm s,\,2}|.
\end{equation}
The initial conditions change here to $c_2=\rho_{33}(0)-\rho_{00}(0)=-2/3$, $\rho_{11}(0) + \rho_{22}(0)=1/3$ and $\rho_{12}(0)=1/6$. The quantum beat is once more present and this time it has full visibility, since the two coefficients in the expansion of Eq.~\eqref{eq:beat2} have equal amplitudes. This is the starting point to determine the intensity correlation function for the side-scattered photons, calculated in Ref.~\cite{Lledo2021}.

For the third case we employ a virtual ``two-photon photoelectric detector'' with jump super-operator $\mathcal{J}\rho=\kappa^{\prime} a^{\dagger 2}\rho a^2$~\cite{Cresser2001} to record a two-photon emission along either of the paths $|\xi_3\rangle \to |\xi_2\rangle \to |\xi_0\rangle$ and $|\xi_3\rangle \to |\xi_1\rangle \to |\xi_0\rangle$. This record conditions the atom-cavity system to its ground state, $\rho_{\rm cond}=|\xi_0\rangle \langle \xi_0|$. Here the quantum beat is absent and the transient response evinces Rabi oscillations with frequency $2\Omega$, decaying at a rate equal to that of spontaneous emission until steady state is reached. 

Figure~\ref{fig:Wigfuncs} focuses on the initial stage of the evolution governed by the master equation for a saturated two-photon resonance with $\Omega/\gamma \approx 5$ ($p_3=0.2475$); for this value of the scaled Rabi frequency in ordinary resonance fluorescence, the optical spectrum would evince a pronounced Mollow triplet~\cite{CarmichaelQO1}. The evolution is conditioned on a single-photon or a two-photon emission at $\tau=0$, in the phase space representation. The three phase space profiles in Figs.~\ref{fig:Wigfuncs}(b)--(d) correspond to three time instants in the course of the conditional evolution, spanning a very short time interval $\gamma \Delta \tau \approx 0.01$ on the order of the period of the quantum beat $\gamma T=2\pi \gamma/(2g + \delta_2-\delta_1)=0.0062$. The differences between them, however, are substantial. The two peaks along the line $y=-x$, observed in all three, are remnants of the steady-state bimodal profile as predicted by Eq.~\eqref{eq:Wss}. We start from an experimentally accessible parameter~\cite{Lutterbach1997, Nogues2000}, the value of the Wigner function at the origin ($x=y=0$), calculated via 
\begin{equation}\label{eq:W0origin}
 W(x=y=0; \tau)=\frac{2}{\pi}\sum_{n=0}^{2} (-1)^n \braket{n|\rho_{c}(\tau)|n},
\end{equation}
and the elements of the density matrix $\rho$ in the four-level model are here taken with respect to the Fock states $\ket{n}$, $n=0, 1, 2$. In Fig.~\ref{fig:Wigfuncs}(a) we follow the evolution of the Wigner function evaluated at the origin for a time interval on the order of the time waited for the loss of the very first photon from an empty cavity with $\braket{a^{\dagger}a}_{\rm ss} \approx 1$ [in fact, for the two-photon resonance (for a cavity filled with a two-level atom) the steady-state photon number is $\braket{a^{\dagger}a}_{\rm ss}=(5/2)p_3$, a quantity which is less than unity for every value of the scaled drive $\varepsilon_d/g$ within the scope of the perturbative treatment]. The Wigner function turns nowhere negative again past that time. Due to the JC interaction establishing a two-photon resonance, however, the non-exclusive probability for emitting an additional photon is maximized earlier, at the delay $\gamma\tau=0.3235$ for the parameters used in the figure -- the time when the intensity correlation function for the forwards scattered light reaches its global maximum.

\begin{figure*}
\begin{center}
\includegraphics[width=\textwidth]{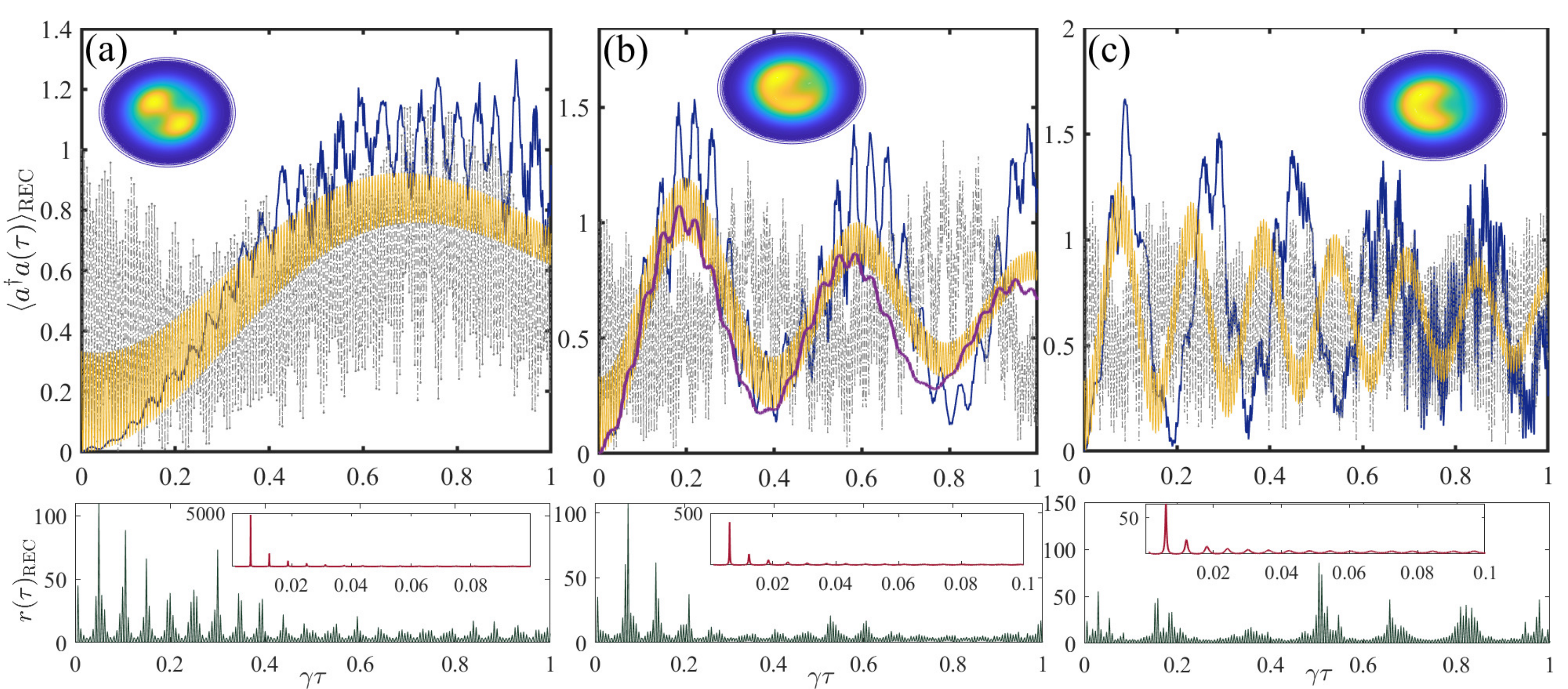}
\end{center}
\caption{{\it Parallel timescales in conditional transients.} Individual realizations unraveling the solution of the ME~\eqref{eq:ME1} in a frame rotating with the drive frequency, extracted from the two-photon resonance operation with $\Delta\omega_d=-g/\sqrt{2}-\sqrt{2}\varepsilon_d^2/g$, initialized to the ground state (thick blue line) and to $|\psi_{\rm s,\,2}\rangle=|1,-\rangle$ (dash-dotted gray line) depicting the conditional average $\braket{a^{\dagger}a(\tau)}_{\rm REC}$ for $\varepsilon_d/g$ equal to {\bf (a)} $0.04$ ($p_3=0.2384$), {\bf (b)} $0.075$ ($p_3=0.2490$) and {\bf (c)} $0.12$ ($p_3=0.2498$). The solid purple line in frame (b) depicts the average photon number obtained from the solution of the ME~\eqref{eq:ME1} (pure-state ensemble average) when the system is initialized to the ground state $\ket{\xi_0}$. The conditional flux ratio $r_{\rm REC}(\tau)$ is depicted below each of the main panels, for the corresponding value of $\varepsilon_d/g$, with the system initialized to $|1,-\rangle$. The trajectories are generated via a quantum state diffusion equation numerically solved by means of a Cash-Karp Runge-Kutta algorithm with adaptive stepsize~\cite{Gisin1992, Schack1997}. The solid yellow lines depict the average photon number extracted from the ME~\eqref{eq:ME2} for the drive amplitudes used in frames (a-c) and with the initial state given by Eq.~\eqref{eq:mixedspon}. The insets of the main panels depict schematic contour plots of the steady-state Wigner function [from Eq.~\eqref{eq:ME1}], while the insets of the minor panels depict the ratio $r(\tau)$ calculated from Eq.~\eqref{eq:ratio4level} starting as well from the state~\eqref{eq:mixedspon}. We take $g/\gamma=500$ and $\gamma=2\kappa$. The depicted results were obtained using a basis truncated at the $N=35$ photon level.}
\label{fig:condtraj}
\end{figure*}

\begin{figure}
\includegraphics[width=0.5\textwidth]{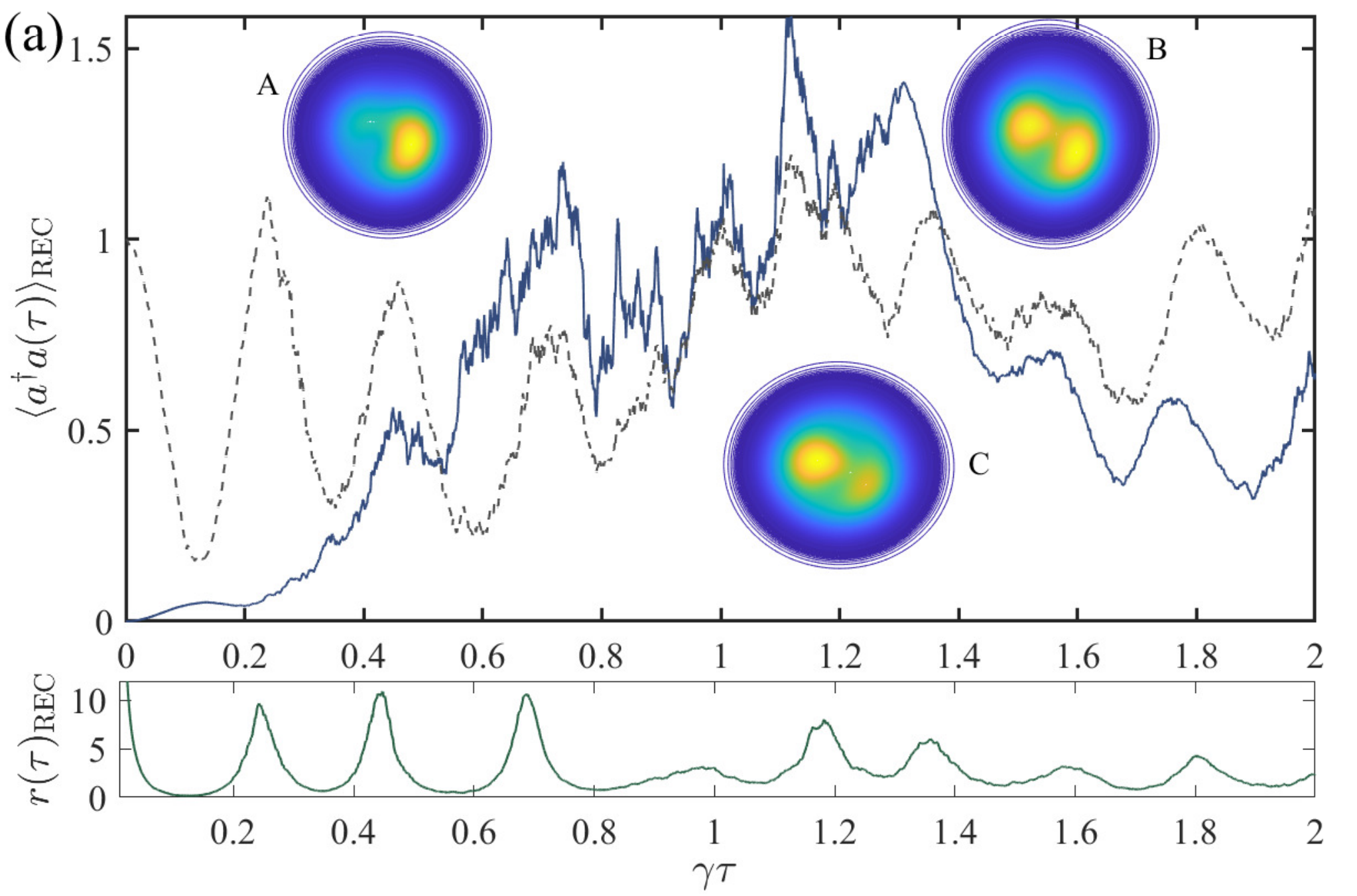}
\includegraphics[width=0.5\textwidth]{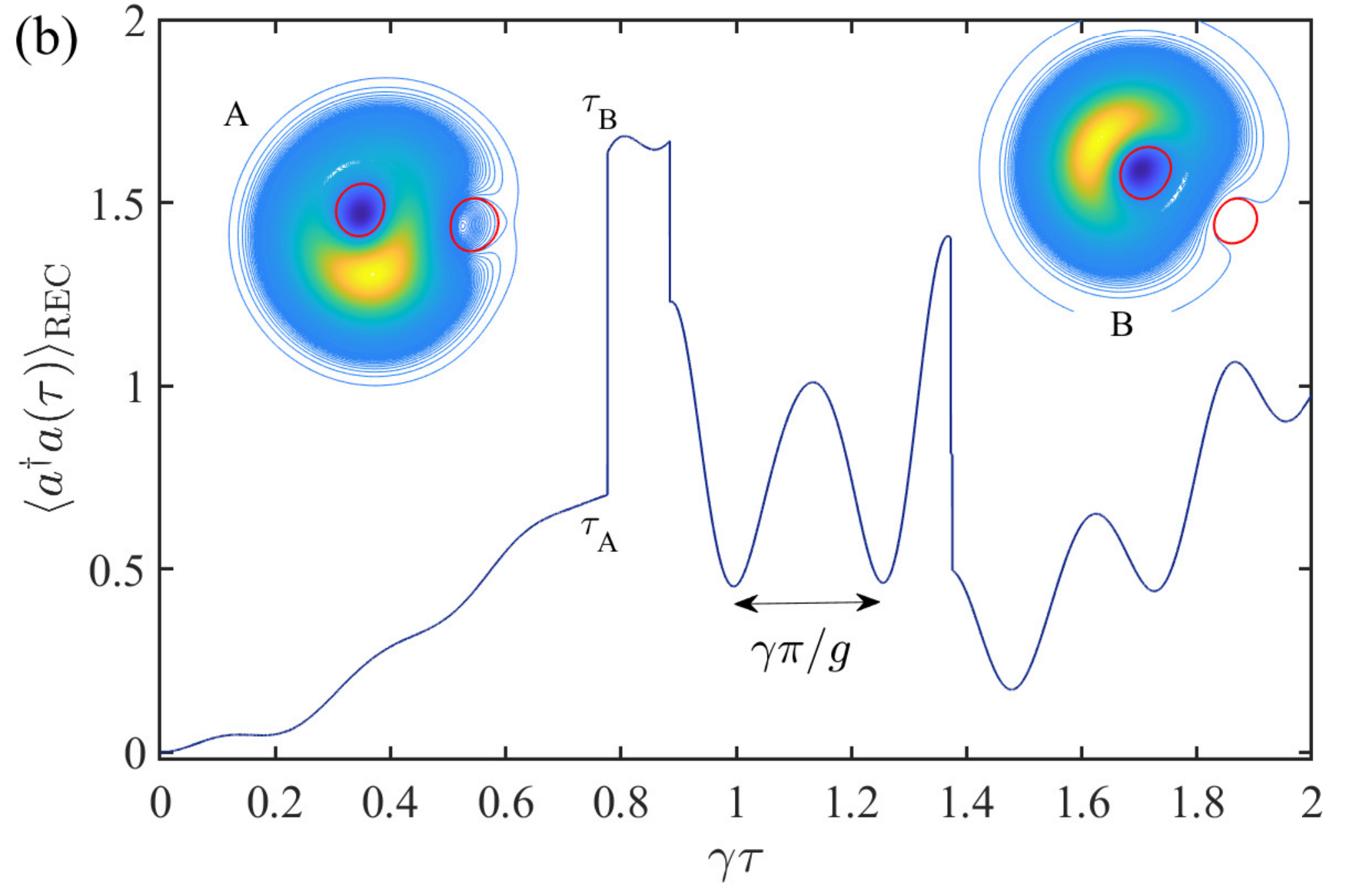}
\caption{{\it Bimodality and photon scattering for lower cooperativity.} {\bf (a)}Sample quantum trajectories unraveling the solution of the ME~\eqref{eq:ME1} (in a frame rotating with the drive) extracted from driving the cavity field at the ``bare'' two-photon resonance with $\Delta\omega_d=-g/\sqrt{2}$, initialized to the ground state (thick blue line) and to $|\psi_{\rm s,\,2}\rangle=|1,-\rangle$ (dashed gray line) depicting the conditional average $\braket{a^{\dagger}a(\tau)}_{\rm REC}$ for $\varepsilon_d/g=0.23$, $g/\gamma=12$ and $\gamma=2\kappa$. The single-atom cooperativity is $\mathcal{C}\equiv g^2/(\kappa \gamma)=288$, corresponding to the strong-coupling parameters attained in the experiment of Ref.~\cite{Najer2019} when extracting photon correlations from a microcavity-quantum dot system (with $\mathcal{C}_{\rm max} \approx 150$). Schematic representations of the Wigner function are given in the three inset contour plots for cooperativity parameters $\mathcal{C}/4$ (in A), $\mathcal{C}$ (in B) and $4\mathcal{C}$ (in C). The steady-state average intracavity photon number $\braket{a^{\dagger}a}_{\rm ss}$ is $0.60$, $0.80$ and $0.87$, respectively. As in Fig.~\ref{fig:condtraj}, the minor panel depicts the conditional flux ratio $r_{\rm REC}(\tau)$ for a system initialized to $|1,-\rangle$. The results presented were obtained using a basis truncated at the $N=35$ photon level. {\bf (b)} Sample quantum trajectory, initialized to the ground state, depicting the conditional average $\braket{a^{\dagger}a(\tau)}_{\rm REC}$ for the same parameters as in (a) [for a single-atom cooperativity $\mathcal{C}=288$]. The scattering record is obtained by a Monte Carlo algorithm using two independent random numbers, each uniformly distributed on the unit line (see, e.g., Ch. 18 of Ref.~\cite{CarmichaelQO2}). The two schematic inset contour plots depict the (instantaneous) Wigner function of the intracavity field at the times $\tau_{A,B}$ corresponding to the start and the end of the first jump, respectively, as indicated in the plot. The red contours demark the regions of negative values.}
\label{fig:Coopl}
\end{figure}

The pronounced quantum beat makes the Wigner function at the origin reach negative values earlier for a photon emitted as fluorescence than in transmission -- the dotted orange curve is the first to cross the zero line. If the Rabi frequency is somewhat smaller, corresponding to $p_3=0.24$ for which there is better agreement with the numerical predictions based on ME~\eqref{eq:ME1}, then the Wigner function at the origin always remains positive following a forwards photon emission at $\tau=0$, while an emission sideways brings it to negative values for a time window $\gamma\Delta \tau \approx 0.25$ centered around $\gamma\tau=0.67$, in coordination with several periods of the quantum beat. Figure~\ref{fig:Wigfuncs}(b) depicts the Wigner distribution in phase space at the time when the expectation of the parity operator for the cavity field [we recall that $W(\alpha=0)=(2/\pi)\,{\rm tr}[\exp(i \pi a^{\dagger}a)\,\rho]$ -- see~\cite{Nogues2000}] is minimized (where it turns most negative), following a single-photon emission at $\tau=0$. This occurs one beat period past the time when the intensity correlation function is maximized (although this is not resolved on the scale of the figure). For an initial two-photon emission, on the other hand, the minimum parity expectation is positive. Nevertheless, the {\it quasi}probability distribution takes negative values inside a ring about the origin, where we also find two dips along the line $y=x$ in conjunction with two extended peaks along $y=-x$. Hence, remnants of steady-state bistability appear very shortly after a (two-)photon emission establishes a circularly symmetric distribution (in the present case a Gaussian centered at the origin). For an initial single-photon emission, the Wigner function coordinates with the quantum beat. This instance is revealed in Fig.~\ref{fig:Wigfuncs}(d) depicting the {\it quasi}probability distribution at the local maximum of $W(x=y=0;\tau)$ following the global minimum; the quantum interference between the two intermediate states, $|\xi_1\rangle$ and $|\xi_2 \rangle$ has now reversed the dip of the single-photon state $\ket{1}$ to a peak about the phase-space origin. We note here that when we excite the three-photon resonance instead, a superposition of quantum beats appears both in the parity and photon correlation, as reported in Ref.~\cite{Mavrogordatos2022}.

So far we have kept our eyes on the transient parity beating of the cavity field, arising as a solution to the effective ME~\eqref{eq:ME2} for the two-photon resonance operation. The effect of the different transitions occurring during the cascade forming a two-photon resonance is revealed by individual quantum trajectories unraveling the solution of the ME~\eqref{eq:ME1}, best illustrated here by focusing on a side scattering event -- a special case for which the quantum beat has full visibility. A spontaneously emitted photon prepares the conditional density matrix of Eq.~\eqref{eq:mixedspon}, which means that $2/3$ of the realizations will be initialized to the vacuum, and the remaining fraction to the superposition state $|\psi_{\rm s,\,2} \rangle=|1,- \rangle$ -- this involves the first excited couplet states. In the latter, the quantum beat (of frequency $\approx 2g$) appears from the very start, as we can observe in all panels of Fig.~\ref{fig:condtraj}, while in the former the beat is gradually superimposed on top of the Rabi oscillation at frequency $2\Omega$. The average photon number obtained from the ME~\eqref{eq:ME2}, on the other hand, directly reveals both these timescales from the very start.

A quantity of significant interest in nonperturbative cavity QED is the ratio of the photon flux emitted along the axis of the cavity to that emitted off the sides~\cite{Carmichael1993, Cui2006, KochM2011, CarmichaelQO2}, together with the photon cross-correlation~\cite{Hughes2011}. When conditioned on a particular scattering record, this ratio is given by the expression $r_{\rm REC}(\tau)=2\kappa\braket{(a^{\dagger}a)(\tau)}_{\rm REC}/(\gamma \braket{(\sigma_{+}\sigma_{-})(\tau)}_{\rm REC})$. According to the prediction of the four-level model, it takes the steady-state value $r_{\rm ss}=(5p_3/2)/(3p_3/2)=5/3$ (for $\gamma=2\kappa$) independent of any conditioning~\cite{Lledo2021, Wigner2PB}. In the transient, it is determined by the time-varying matrix elements whose evolution is governed by the effective ME~\eqref{eq:ME2}, as
\begin{equation}\label{eq:ratio4level}
 r(\tau)\equiv \frac{2\kappa\, {\rm tr}[\rho(\tau) a^{\dagger}a]}{\gamma\, {\rm tr}[\rho(\tau)\sigma_{+}\sigma_{-}]}=\frac{\rho_{11} + \rho_{22} + 3\rho_{33} + 2{\rm Re}(\rho_{12})}{\rho_{11} + \rho_{22} + \rho_{33} - 2{\rm Re}(\rho_{12})},
\end{equation}
where the initial condition is set by the mixed state of Eq.~\eqref{eq:mixedspon} following an emission of a side-scattered photon. The three plots in the minor panels of Fig.~\ref{fig:condtraj} testify to a notable divergence from the steady-state value of the flux ratio when the system state is initialized to $|1,- \rangle$: occasional spike clusters occur, with a periodicity dictated by the intermediate timescale $\approx \gamma/[(1-1/\sqrt{2})g]$ associated with the dominant rate $\Gamma_{31}$ in the cascaded decay [Eq.~\eqref{eq:Gamma31}] -- this oscillation survives in the pure-state ensemble average over trajectories [see Fig.~\ref{fig:condtraj}(b)] but is not captured in the secular approximation and the adiabatic elimination of intermediate states. As the transition saturates and steady-state bimodality collapses, these clusters move further into the transient. Most of these intricacies are missing in the dynamics predicted by the ensemble-averaged flux ratio as given by the solution of the effective master equation: we only observe monotonically decaying peaks with the period of the quantum beat in a considerably smaller fraction of the transient. Quantum jumps interrupt and reinitiate the quantum beat at random times, dephasing the oscillation in the pure-state ensemble average. The peak heights decrease as the two-photon transition saturates and we move away from the realm of the perturbative treatment. Sample trajectories also show that tuning the drive field to excite a vacuum Rabi resonance ($\Delta\omega_d=-g$) and starting from $|1,- \rangle$ produces once more a modulation at frequency $2g$ due to the weak activation of the fist excited couplet states [see Fig. 5 of Ref.~\cite{Tian1992} and Fig. 4(c) of Ref.~\cite{Najer2019}], but now with a larger dynamic Stark splitting which is linear in $\varepsilon_d/g$ and a flux ratio fluctuating about unity past the transient. A steady-state flux ratio equal to unity finds here its explanation in the two-state approximation, for which the fields scattered into both output channels are proportional to one and the same operator, one satisfying the ME of ordinary resonance fluorescence (see Sec. 13.3.3 of Ref.~\cite{CarmichaelQO2}). For the two-photon resonance this is no longer the case.

Let us finally see what a heterodyne detection measurement of the field transmitted from the cavity can reveal about the evolution conditioned on a single emission event, based on the parameters used in the cavity QED experiments of Refs.~\cite{Najer2019} and~\cite{Hamsen2017}. Here the single-atom cooperativity $\mathcal{C} \equiv g^2/(\kappa\gamma)$ can be in excess of $10^2$, yet several orders of magnitude smaller than what was used in Ref.~\cite{Bishop2009} to picture the extended JC spectrum. Our focus remains on the regime of steady-state bimodality, where the intracavity photon number is below unity. The trajectories of Fig.~\ref{fig:Coopl}(a) evince pronounced oscillations at frequency $\approx 2g$ on significantly longer timescale against the coherence time compared to those depicted in Fig.~\ref{fig:condtraj}, interrupted by dynamical bistable switching. The conditional flux ratio once more reflects the first excited couplet state splitting, although here the variation is much smoother. As far as the solution of the ME~\eqref{eq:ME1} in the rotating frame is concerned, for a fixed ratio of $\varepsilon_d/g=0.23$, we find that the average photon number increases by a very small amount and remains below unity with a 16-fold increase in the cooperativity, a variation reflected in the sequential occupation of the two states of amplitude bimodality [see the three insets of Fig.~\ref{fig:Coopl}(a)]. 

The oscillations due to the splitting of the first couplet states are more clearly visible for an alternative unraveling of the ME solution, one corresponding to direct photodetection down the two channels (cavity decay and spontaneous emission). The pattern is however interrupted by sudden jumps conforming to the underlying bistability, as can be seen in Fig.~\ref{fig:Coopl}(b). The two Wigner functions of the cavity field, resolving the first such discontinuity, show a squeezed state swirling around a central region of negative values -- the imprint of a variable interference the single-photon state $\ket{1}$ -- when about one quantum is added to the conditional photon number [an increase of $\braket{a^{\dagger}a(\tau)}_{\rm REC}$ from $0.70$ to $1.64$ is precipitated by a detectable scattering event in the forwards direction]. This instance provides evidence for the dynamical participation of {\it countable photons} in the cavity state, in the course of individual scattering records during the transient -- despite the fact that the steady-state distribution is everywhere positive -- a property which also manifests itself in the negativity of the ensemble average distributions of the intracavity field (see Fig.~\ref{fig:Wigfuncs}).

In conclusion, we recall that from the point of view of quantum optics, an open system is approached as a generalized scattering center, in other words the mediator of a transformation (scattering) from input fields to output fields, with inputs and outputs treated as excitations of a reservoir or bath~\cite{OpenQuantCh4}. Ultimately, when the dynamical response is Markovian, outputs may be measured without disturbing the unconditional open system evolution. More specifically, in cavity QED and its ramifications, our goal is to redistribute the photons in the incoming beam: the interaction of the two-level atom with a first photon must therefore change the probability for the transmission of the next~\cite{CarmichaelTalk2000}. Taking that route, in this Letter we have operationally distinguished forwards emission from the photons scattered off the sides of a cavity for a fundamental scattering configuration of nonperturbative cavity QED, namely the two-photon resonance of the open driven JC model in its strong-coupling limit. The idea of appealing to state reduction dates back to the exemplary experiment of cavity QED realized by Foster and coworkers (resonant excitation of a few atoms), measuring a wave-particle correlation function which recorded the conditional time evolution of the optical field -- equilibrium was there approached via the vacuum Rabi oscillation of a fraction of a photon~\cite{Foster2000A}. Poor collection efficiencies and the dead-time of photon counters in the experiments of atomic physics often impede the faithful attribution of each individual recorded click to their origin, an obstacle which can be overcome in circuit QED~\cite{Katz2008, Minev2019}. Tracking the resulting conditional evolution is not however merely a book-keeping matter: different transitions trigger a distinct quantum interference between dressed states, which is reflected in the intracavity field and its moments. With the exception of a two-photon jump, all other emissions leave the system in a mixed state, dictating the proportion of individual realizations that are to display certain dynamical aspects directly associated with the JC spectrum. The aforementioned interference is intertwined with the semiclassical Rabi oscillation accompanying a saturable transition, and the result is read in the correlation between the forwards or side-scattered photons.

Any possible emission event occurring after the attainment of steady state destroys the bimodal phase-space profile of the cavity field state and imposes an azimuthal symmetry to the Wigner distribution. In this Letter, we have followed the evolution of system density matrix, working against the initial conditional symmetry, as bimodality is gradually restored in the form of a weighted average over dressed states (detailed balance) together with an off-diagonal symmetry-breaking term imposed by the two-photon drive [see Eq.~\eqref{eq:Wss}]. When the two-photon transition saturates, the intensity correlation function of the forwards-scattered light falls below unity at zero delay~\cite{Shamailov2010}, which means that the simultaneous emission of two photons is increasingly unlikely. Consequently, a quantum beat is more likely to appear after any perturbation of the steady state when the latter is markedly bimodal,  compared to a distribution resembling that of the vacuum state. This will also entail a modulation of the flux ratio at a frequency $\approx 2g$, with the forwards dominating over the side flux in the steady state, while the two are equal when exciting a vacuum Rabi resonance. As cavity QED experiments move to higher (single-atom) cooperativity parameters, not only will the occurrence of a pronounced quantum beat in single realizations indicate the conditioning on a previous emission, but one may readily explore the transition from bunching to antibunching (varying the drive amplitude) through the {\it quasi}probability distribution of the cavity field (at present, the quantum beat has been demonstrated only in the presence of photon bunching in the strong-coupling regime -- see Figs. 3 and 4 of Ref.~\cite{Najer2019}). Quantum trajectories simulating single heterodyne detection records~\cite{CarmichaelQO2} reveal a disparity between the forwards and sideways emitted photon flux, correlated with the quantum beat, the rates of the cascaded decay and the semiclassical ringing [also compare with Fig. 1(a) of Ref.~\cite{KochM2011}]. On the other hand, the probability distribution of the net charge deposited in a detector in temporally mode-matched homodyne detection measures a marginal of the Wigner distribution representing the state of the intracavity field immediately before the free decay (see~\cite{Cohstates94} and Ch. 18 of Ref.~\cite{CarmichaelQO2}). Future work, motivated by Fig.~\ref{fig:Coopl}(b), should then address the link between individual realizations obtained for different initial wavefunctions as {\it observed} scattering records and the {\it quasi}probability distribution of the evolving conditional electromagnetic-field state.  

\begin{acknowledgments}
I wish to thank C. Lled\'{o} for his help with the numerical results displayed in Fig. 3(b). I acknowledge financial support by the Swedish Research Council (VR) in conjunction with the Knut and Alice Wallenberg foundation (KAW). This work was supported by the Agencia Estatal de Investigaci\'{o}n (the R\&D project CEX2019-000910-S, funded by MCIN/ AEI/10.13039/501100011033, Plan National FIDEUA PID2019-106901GB-I00, FPI), the Fundaci\'{o} Privada Cellex, Fundaci\'{o} Mir-Puig, and by the Generalitat de Catalunya (AGAUR Grant No. 2017 SGR 1341, CERCA program). 
\end{acknowledgments}

\bibliography{bibliography}
\end{document}